\documentstyle[12pt]{article}

\newcommand{\secn}[1]{Section~\ref{#1}}
\def\O{{\cal O}}
\def\L{{\cal L}}
\newcommand{\bra}[1]{\langle{#1}|}
\newcommand{\ket}[1]{|{#1}\rangle}

\newcommand{\tbl}[1]{Table~\ref{#1}}
\newcommand{\eq}[1]{Eq.~(\ref{#1})}

\newcommand{\nl}{\nonumber \\}

\newcommand{\sigmav}{\mbox{\boldmath$\sigma$}}
\newcommand{\gammav}{\mbox{\boldmath$\gamma$}}
\newcommand{\Dv}{{\bf D}}
\newcommand{\Ev}{{\bf E}}
\newcommand{\Av}{{\bf A}}
\newcommand{\Bv}{{\bf B}}
\newcommand{\Cv}{{\bf C}}
\newcommand{\pv}{{\bf p}}
\newcommand{\kv}{{\bf k}}

\newcommand{\qv}{{\bf q}}
\newcommand{\delv}{\mbox{\boldmath$\nabla$}}

\newcommand{\psid}{{\psi^\dagger}}

\newcommand{\dplus}[1]{{\Delta^{(+)}_{#1}}}
\newcommand{\dminus}[1]{{\Delta^{(-)}_{#1}}}
\newcommand{\dpm}[1]{{\Delta^{(\pm)}_{#1}}}
\newcommand{\dsq}{{{\bf\Delta}^{(2)}}}

\newcommand{\R}{{\cal R}}
\newcommand{\F}[2]{{F_{#1 #2}}}
\newcommand{\Fc}[2]{{F^{(c)}_{#1 #2}}}
\newcommand{\U}[2]{{U_{#1,#2}}}
\newcommand{\Udag}[2]{{U^\dagger_{#1,#2}}}
\def\eff{{\rm eff}}
\def\e{{\rm e}}
\def\dH{{\delta H}}

\newcommand{\Deltavpm}{{\bf\Delta^{(\pm)}}}
\newcommand{\Deltavtpm}{{\tilde{\bf\Delta}^{(\pm)}}}
\newcommand{\xv}{{\bf x}}

\def\beq{\begin{equation}}
\def\eeq{\end{equation}}
\def\beqa{\begin{eqnarray}}
\def\eeqa{\end{eqnarray}}
\def\chid{{\chi^\dagger}}
\def\etad{{\eta^\dagger}}

\begin{document}

\rightline{CLNS 92/1136}
\rightline{OHSTPY-HEP-T-92-001}
\rightline{\hfill February, 1992}

\vskip 1cm

\centerline{\Large \bf Improved Nonrelativistic QCD}
\centerline{\Large \bf for Heavy Quark Physics}

\vskip 1cm

\centerline{\large G. Peter Lepage,
                   \footnote{e-mail: gpl@lnssun1.tn.cornell.edu} \
            Lorenzo Magnea
                   \footnote{On leave from Universit\`a di Torino, Italy} \
            and Charles Nakhleh }
\centerline{Newman Laboratory of Nuclear Studies}
\centerline{Cornell University, Ithaca, NY 14853}

\vskip .5cm

\centerline{\large Ulrika Magnea}
\centerline{State University of New York at Stony Brook}
\centerline{Stony Brook, NY 11794}

\vskip .5cm

\centerline{\large Kent Hornbostel}
\centerline{The Ohio State University}
\centerline{Columbus, OH 43210}

\vskip 1.5cm

\begin{abstract}

We construct an improved version of nonrelativistic QCD for use in lattice
simulations of heavy quark physics, with
the goal of reducing systematic errors from
all sources to below 10\%. We develop power counting rules to assess the
importance of the various operators in the action and compute
all leading order corrections required by relativity and finite lattice
spacing. We discuss radiative corrections to tree level coupling
constants, presenting a procedure that effectively resums the largest such
corrections to all orders in perturbation theory. Finally, we comment on
the size of nonperturbative contributions to the coupling constants.

\end{abstract}

\newpage

\section{Introduction}

\label{intro}

Several papers in recent years have explored the possibility of substituting
a nonrelativistic action for the usual Dirac action in lattice QCD simulations
of heavy quarks \cite{leptacone,leptactwo,davtac} .
In particular, this technique has been applied to
simulations of the $\psi$ and $\Upsilon$ families of mesons. Results
from these simulations have been very encouraging. They not only
capture the gross features of quarkonium physics but also are surprisingly
accurate given the many approximations inherent in the simulations. Such
success follows from a variety of properties unique to quarkonium states and
from the algorithmic advantages of a nonrelativistic formulation of quark
dynamics. This suggests the possibility of a truly reliable numerical study of
quarkonium states using computing resources that are currently available, a
study in which all systematic errors are identified, computed, and removed to
some reasonable level of precision. In this paper we develop the outline for
such a program and provide some of the ingredients needed to move from
existing quarkonium studies to precision studies. Most of the techniques
we describe are readily generalized for use in simulations of other
bound states containing heavy quarks, such as $D$ and $B$ mesons.

The major tools for simulating gluon dynamics in QCD all involve Monte Carlo
methods of one sort or another. Consequently, in simulating quarkonium
states, we must worry about statistical as well as systematic
errors. However, simulation data can be generated far more efficiently for
quarkonium states than for ordinary hadrons, and thus the dominant
errors for many of the most important quarkonium measurements are systematic.
Here we focus only on these.

The nonrelativistic formulation of heavy-quark dynamics is referred to as
nonrelativistic quantum chromodynamics or NRQCD. There are five
potentially significant sources of systematic error when NRQCD is used to
study quarkonium states:
 \begin{itemize}
 \item {\em Relativity.\/} The basic action for the
heavy quarks incorporates the physics of the Schr\"{o}dinger equation.
Left out are relativistic corrections of $\O(v^2)$, where $v$ is the quark
velocity. Since $v^2$ is approximately  0.3 for the $\psi$'s and 0.1 for the
$\Upsilon$'s, we expect errors of 10--30\% due to relativistic
corrections. Such errors can be systematically removed by adding new
interactions to the  heavy-quark action.
 \item {\em Finite Lattice Spacing.\/} As in all lattice simulations, there
are errors resulting from the discretization of space and time. For our
systems, such errors are $\O(a^2p^2)$ and $\O(aK)$, where $a$ is the lattice
spacing and $p$ and $K$ are the typical three-momentum and kinetic
energy of a quark in the hadron. Since typically $a\sim 1/M$, where $M$ is
the quark mass, these errors are roughly the same order as those due to
relativity.  These may also be systematically removed by correcting the
heavy-quark action.
 \item {\em Radiative Corrections.\/} As in the Dirac action, the basic action
in NRQCD contains only two parameters: the QCD coupling constant~$g$, and
the quark mass~$M$.\footnote{
Actually there is a third parameter, $E_0$, which adjusts the
origin of the nonrelativistic energy scale to coincide with that of the
relativistic energy scale. However, this parameter affects only the absolute
masses of hadrons and has no effect upon nonrelativistic quark dynamics,
mass splittings, wavefunctions, etc. In principle $E_0$ may be computed, at
least
approximately, using perturbation theory.}
However, each of the new interactions added to NRQCD to correct for
relativity or finite lattice spacing introduces a new coupling constant.
These new couplings can be treated as free parameters, to be adjusted by
fitting to experimental data, but this significantly reduces the predictive
power
of  a simulation. Alternatively one can try to compute these couplings in
terms of $g$ and $M$ using perturbation theory, thereby reducing the number
of parameters back to two. The radiative corrections to these coupling
constants are determined by physics at distances of order the lattice
spacing~$a$, and therefore perturbation theory should be applicable provided
$a$ is small enough. However, $a$ must not be too small. NRQCD is
nonrenormalizable, and as a result the radiative corrections contain
power-law ultraviolet divergences, like $g^2/aM$, which make perturbation
theory
useless unless $a\sim 1/M$ or larger. Lattice spacings currently in use,
where $g^2\approx 1$, are probably suitable for both $c$ and $b$ quarks, and
also
seem small enough for perturbation theory to work. Radiative corrections of
order 20--40\% could easily result, making these among the largest
contributions at the order we are considering. We will see that they are
well estimated using mean-field theory techniques.
Uncalculable corrections due to nonperturbative physics at short
distances could be of order a few percent.
 \item {\em Finite Lattice Volume.\/} Quarkonium hadrons are much smaller
than ordinary hadrons. Typical lattices in use today can easily be 10
$\Upsilon$-radii across, making it unlikely that the finite volume of
the lattice has much of an effect upon simulation results.
 \item {\em Light-Quark Vacuum Polarization.\/} Hadrons built of heavy quarks
are affected by light quarks through vacuum polarization. Such effects are
usually very expensive to simulate numerically due to the
inefficiency of the algorithms when quark masses are small. However, effects
from light-quark vacuum polarization are likely to be small for
quarkonium states. Some evidence comes from the decay widths of excited
$\psi$ and $\Upsilon$ mesons into $D$ and $B$ mesons, respectively. These
decays result from light-quark vacuum polarization, and their widths are
closely related to the energy shifts due to light quarks. The small size of
these widths compared to typical mass splittings, together with
model studies of this coupled-channel problem \cite{eich} ,
suggests that light-quark
vacuum polarization might affect quarkonium states at only the 10\%
level. Furthermore, the extrapolation from large to small light-quark
masses should be quite smooth for low-lying quarkonium states since, unlike
the $\rho$ meson, for example, these are all effectively stable even for
realistic
light-quark masses.\footnote{ A low-lying quarkonium state does decay, via
the annihilation of the heavy quarks into gluons. The decay rate is
relatively small and the decay mechanism has a negligible effect upon other
properties of the hadron. Consequently, this decay mechanism is usually
omitted in simulations.} Note finally that heavy-quark vacuum polarization
has only a small effect upon quarkonium states since there is too little
energy in such states to easily produce a heavy virtual quark and antiquark.
For most purposes such effects can be ignored, though they are easily
incorporated through perturbatively computed corrections to the gluon
action.
 \end{itemize}

This list suggests that we might be able to reduce
systematic errors below 10\% in quarkonium simulations using NRQCD. To
achieve this goal we must compute all of the leading-order interactions
that correct the action for relativity and finite lattice spacing as
well as the leading radiative corrections for the most important of these
interactions. Other sources of error are probably unimportant at this level,
although including some light-quark vacuum polarization effects (by
extrapolation) is both feasible and desirable. In this paper, we present all
tree-level corrections to the NRQCD action that are relevant to work at this
level of precision. Although we have only just started the analysis of
radiative corrections, we also present here some general comments concerning
the nature and treatment of these corrections.

Our analysis begins in \secn{pcounting} with the development of
power-counting rules that allow us to determine the relative importance of
various operators in the NRQCD action for analyses of quarkonium states.
Such rules are trivial when one is considering light-heavy systems such as $D$
or $B$~mesons. They are more complicated for heavy-heavy systems since the
heavy quark is not a spectator in such a hadron; in particular, the quark
mass strongly affects the hadron's size and internal dynamics.
In \secn{rel}, we compute the $\O(v^2)$ corrections to the NRQCD action in the
continuum for both Minkowskian and Euclidean space. In \secn{latt}, we
discretize the corrected NRQCD action, further modifying it to remove the
leading errors due to the finite lattice spacing. In \secn{alpha}, we discuss
some general properties of the radiative corrections, we
give estimates for the largest of these, and we discuss the extent to which
nonperturbative physics can affect the NRQCD couplings.
Finally, in \secn{concl}, we summarize our results and discuss future
prospects.

\section{Power-Counting Rules for NRQCD}

\label{pcounting}

NRQCD is an effective field theory which approximates ordinary relativistic
QCD at low energies.
In principle, the NRQCD action may be corrected to
reproduce the exact results of QCD by including an infinite number of
nonrenormalizable interactions. In practice, however,
exact agreement is unnecessary: only a finite number of
interactions are required to attain any desired accuracy.
Some criterion is necessary to
assess the importance of various interactions when designing the corrected
action. In this section we categorize interactions by
their effect on the energy spectrum of quarkonium states.

\subsection{Building Blocks for NRQCD}

Correction terms $\delta\L(x)$ in the lagrangian density of NRQCD contribute
 \begin{equation} \label{dE}
 \delta E = - \bra{n}\int d^3x\,\delta\L(x) \ket{n}
 \end{equation}
to the energy of a quarkonium state $\ket{n}$. Since operators are not
normal-ordered in lattice simulations, there are cases where large
contributions to $\delta E$ involve contractions of two or more fields in
a single term of $\delta\L(x)$.
Such contributions have no bearing upon the relative
importance of corrections since their effect is
cancelled by shifts in the coupling constants of
lower-dimension interactions. Thus we may ignore
these contributions when estimating the importance of
various terms. For the remaining contributions,  each  field and
operator in $\delta\L(x)$ has a characteristic magnitude determined
by the dynamics of the hadron.

The important dynamical scales for the
nonrelativistic analysis of the $\psi$ and $\Upsilon$
states are the typical momenta~$p$ and
kinetic energies~$K$ of their quarks; these are of order
1--2~GeV and 500~MeV respectively.  If $v$
is the quark velocity, the two scales are related to the
quark mass~$M$ by
 \begin{equation}
 p\sim M v \qquad K \sim M v^2.
 \end{equation}
These determine the characteristic magnitudes of the fundamental fields
and operators in NRQCD and allow us to estimate the (normal-ordered)
contribution of any interaction to a quarkonium energy.

The continuum action for NRQCD is built from the fields and operators shown in
\tbl{fields}. We also list the estimated magnitude for
each of these in terms of $v$ and $M$.
It is important to remember that it is the order in $v$
rather than the dimension of an operator that determines
its numerical importance.
In this table, the quark field~$\psi$
and antiquark field~$\chi$ have two spin components and three
color components that transform in the fundamental representation.
Combinations such as
$\psi^\dagger\psi$ or $\chi^\dagger\psi$ are invariant under rotations
and color transformations, while
$\psi^\dagger\sigmav\psi$, for example, is a color singlet but
rotates as a vector.
The gauge fields have two color indices and transform in the adjoint
representation.  For example, the chromoelectric field is
 \begin{equation}
 \Ev(x) \equiv \Ev^a(x)\,T^a .
 \end{equation}
The generators $T^a$ are normalized such that
${\rm Tr}(T^a T^b) = \delta^{ab}/2$.
The combination $\psi^\dagger\Ev \psi$
is then a color-singlet vector field.  Covariant derivatives act
on the quark and gauge fields according to
 \begin{eqnarray}
 D_t \psi(x) &\equiv& \left(\partial_t +ig\phi(x)\right) \psi(x) \nonumber \\
 \Dv \psi(x) &\equiv& \left(\delv -ig\Av(x)\right) \psi(x) \nonumber \\
 D_t E_i(x) &\equiv& \partial_t E_i(x) + i[g\phi(x),E_i(x)] \\
 \Dv E_i(x) &\equiv& \delv E_i(x) -i[g\Av(x),E_i(x)] \, , \nonumber
 \end{eqnarray}
where $\phi$ and $\Av$ are the scalar and vector gauge potentials, and
$g$ is the coupling constant. The last definitions insure that the chain
rule works properly for covariant derivatives.  For example,
 \begin{equation}
 D(\Ev \psi) = (D\Ev) \psi + \Ev (D\psi).
 \end{equation}
The nonabelian electric and magnetic fields can be defined in
terms of these covariant derivatives:
 \begin{eqnarray}
 \label{ebcomm}
 [ D_t,\Dv ]\,\psi &\equiv & ig\Ev \psi \nl
 \mbox{} [ D_i,D_j ]\,\psi &\equiv & -i g \epsilon_{ijk} B^k \psi .
 \end{eqnarray}

The estimated sizes given in \tbl{fields} follow from the properties
of quarkonium states. For example, the number operator for
(heavy) quarks,
 \begin{equation}
 \int d^3x\, \psi^\dagger(x)\,\psi(x),
 \end{equation}
has an expectation value very near 1 for a quarkonium meson. Since the quark
in the meson is localized within a region $\Delta x \sim 1/p$, we
estimate
 \begin{equation}
 \int d^3x \sim \frac{1}{p^3}\, ,
 \end{equation}
and therefore
 \begin{equation}
 \psi^\dagger(x)\,\psi(x) \sim p^3 \, ,
 \end{equation}
that is, $\psi\sim p^{3/2}$.
Similarly, the operator for kinetic energy,
 \begin{equation}
 \int d^3x\,\psi^\dagger(x)\,\frac{\Dv^2}{2M}\, \psi(x)\, ,
 \end{equation}
has an expectation value of $K$ by definition, and so the spatial
covariant derivative acting on a quark field is of order
 \begin{equation}
 \Dv \sim (2\,M\,K)^{1/2} \sim p\, ,
 \end{equation}
as expected.

Field equations can also be used to relate estimates for different operators.
For example, the lowest-order field equation for the quark field,
 \begin{equation}
 \left( iD_t  +\frac{\Dv^2}{2M} \right) \psi = 0\, ,
 \end{equation}
implies that
 \begin{equation}
 D_t \sim \frac{\Dv^2}{2M} \sim K\, ,
 \end{equation}
again as expected. If we specialize to Coulomb gauge, the gauge most natural
for nonrelativistic systems, this equation becomes
 \begin{equation}
 \left( i\partial_t  -g\phi(x) + \frac{\delv^2}{2M} \right) \psi
 \approx 0\, ,
 \end{equation}
where we are neglecting the vector potential, which is small in this gauge,
as discussed below. The potential energy that balances the kinetic energy
and produces a
bound system enters through the operator $g\phi$ and we expect that
 \begin{equation}
 g\phi(x) \sim K
 \end{equation}
in Coulomb gauge.

This result can be checked against the
field equation for $\phi$ (again neglecting the vector potential),
 \begin{equation}
 \delv^2 g\phi(x)  = -g^2\psi^\dagger(x)\psi(x) \, ,
 \end{equation}
which implies that
 \begin{equation}
 g\phi(x) \sim \frac{1}{p^2} g^2 p^3 \sim g^2 p \, .
 \end{equation}
This is consistent with the previous estimate if the effective
low-energy coupling strength is
 \begin{equation}
 \alpha_s \sim g^2 \sim v \, .
 \end{equation}

That the Coulomb-gauge vector potential in a
quarkonium state is typically smaller than
the scalar potential may be inferred from the
field equation for the vector potential,
 \begin{equation}
 \left(\partial_t^2 -\delv^2\right)\,g\Av =
 \frac{g^2}{M}\psi^\dagger\delv\psi + g \phi\delv g\phi +
 \cdots \, .
 \end{equation}
Using the estimates given above, we see that
 \begin{equation}
 g\Av(x) \sim \frac{1}{p^2} \left(\frac{g^2}{M}\,p^4 + p K^2\right)
 \sim v K
 \end{equation}
which is smaller than the scalar potential by a factor of the quark velocity.
Estimates for the nonabelian electric and magnetic fields follow
immediately from
 \begin{eqnarray}
 g\Ev &=& -\delv g\phi + \cdots \,\sim \, p K \nl
 g\Bv &=& \delv\times g\Av + \cdots \,\sim \, K^2\, .
 \end{eqnarray}
In quarkonium, as expected for a nonrelativistic system,
magnetic fields are smaller than electric fields by a factor of $v$.

These estimates are valid perturbatively,
as in the analysis of positronium with NRQED,
but are also consistent nonperturbatively. The essential
ingredient in their derivations is the nonrelativistic nature of quark
dynamics.

\subsection{NRQCD interactions in the continuum}

Equipped with the power-counting rules of the last section, we can enumerate
terms for the quark action in NRQCD that are potentially important for
quarkonium physics.
The leading terms are those of the Schr\"odinger theory:
 \begin{equation} \label{leading}
 S_0 = \int d^4x\,\psi^\dagger(x) \left(iD_t + \frac{\Dv^2}{2M}\right)
 \psi(x).
 \end{equation}
Correction terms due to relativity and finite lattice spacing must respect
the symmetries of the theory: gauge invariance, parity,
rotational symmetry, unitarity, and so on.
For example, the interaction $\psi^\dagger\Ev
\cdot\sigmav\psi$, corresponding to an intrinsic electric dipole moment,
is not allowed because it is odd under parity, while
$\psi^\dagger\Bv\cdot\sigmav\psi$ is even under parity and so is
allowed.
Charge-conjugation invariance requires that the total action be invariant
under the interchange $\psi\leftrightarrow\chi$. Thus the antiquark action
can be obtained directly from the quark action.

Correction terms must also be local. This, combined with the fact
that our theory needs only to be accurate to 10\%, severely restricts the
number of interactions. For example, the interaction~$\psi^\dagger
\Bv^2\psi/M^3$ is consistent with required symmetries and so must appear in
the NRQCD action.  But it is unnecessary for our calculations because it is
suppressed by $v^6$ ($\approx 10^{-3}$ for $\Upsilon$'s) relative to the
leading terms in the action.

Ignoring spin splittings for the moment, only corrections that are suppressed
by $v^2$ relative to the leading terms are needed to achieve accuracy to 10\%,
at
least for $\Upsilon$'s. The only such terms bilinear in the quark field are
the order $v^2$ corrections
 \begin{eqnarray} \label{bilinear}
 \delta\L_{\rm bilinear} &\equiv&
     c_1\,\frac{1}{M^3} \, \psid\Dv^4\psi  \nl
 &+& c_2\,\frac{g}{M^2}\, \psid (\Dv\cdot\Ev - \Ev\cdot\Dv) \psi  \nl
 &+& c_3\,\frac{ig}{M^2}\,\psid\sigmav\cdot(\Dv\times\Ev -
     \Ev\times\Dv)\psi  \\
 &+& c_4\,\frac{g}{M}\, \psid\sigmav\cdot\Bv\psi \, . \nonumber
 \end{eqnarray}
In this and subsequent equations, $\Dv$ acts on all fields to its right.
The dimensionless coefficients~$c_1\ldots c_4$ are functions of the
running coupling constant $\alpha_s(\pi/a)$ and $a M$.

Note that terms involving time derivatives of the quark field, such as
$\psi^\dagger D_t^2 \psi$, are not included. Such
terms greatly complicate the numerical evaluation of quark propagators.
We may avoid them here by suitably redefining the
quark field so that factors of $iD_t$ are in effect replaced by factors of
$-\Dv^2/2M$, in accordance with the field equation for $\psi$,
 \begin{equation}
 iD_t \psi(x) \approx \frac{-\Dv^2}{2M}\,\psi(x) \, .
 \end{equation}
These transformations do not affect masses, on-shell scattering amplitudes,
or other physical quantities.

In addition to the bilinear terms, there are four-fermion contact
interactions involving a quark and antiquark:
 \begin{eqnarray} \label{contact}
 \delta\L_{\rm contact} &\equiv&
 d_1\,\frac{1}{M^2}\, \psi^\dagger\chi\,\chi^\dagger\psi \nl
 &+& d_2\,\frac{1}{M^2}
 \psi^\dagger\sigmav\chi\cdot\chi^\dagger\sigmav\psi\, .
 \end{eqnarray}
These appear to be down by only a single power of $v$. However, similar
interactions do not occur in relativistic continuum QCD, and therefore
such terms can occur here only to one-loop order and beyond. Thus the
coefficients~$d_1$ and~$d_2$ are both of order $\alpha_s^2(\pi/a)$,
making these contact interactions significantly less important than the
bilinear interactions considered above. Contact interactions also occur
between pairs of quarks, but these can only affect heavy-quark baryons.
Four-fermion operators can also couple to colored states:
 \begin{eqnarray} \label{contacolor}
 \delta\L_{\rm color} &\equiv&
 d_3 \,\frac{1}{M^2}\, \sum_a \psi^\dagger T^a \chi\,\chi^\dagger T^a \psi \nl
 &+& d_4 \,\frac{1}{M^2}
 \sum_a  \psi^\dagger T^a \sigmav \chi \cdot \chi^\dagger T^a \sigmav \psi \, ,
 \end{eqnarray}
where~$d_3$ and~$d_4$ are of order $\alpha_s^2(\pi/a)$.
Color-singlet mesons are affected by these interactions, since the meson
can become colored by emitting a virtual gluon. Gluon emission however
is suppressed by $v^2$, so these interactions are again not as important
as the bilinear terms in \eq{bilinear}.

Nontrivial spin dependence first appears in the bilinear correction terms
of \eq{bilinear}, so that spin splittings in quarkonium meson families are
suppressed by $v^2$.  A 10\% determination of these
splittings therefore requires the retention of spin-dependent
interactions suppressed by $v^4$. The additional quark-bilinear
terms required to this order are of three sorts: permutations of
the operators $\psid\sigmav\psi$, $\Bv$, and two $\Dv$'s; permutations of
the operators $\psid\sigmav\psi$, $\Dv\times\Ev$, and two $\Dv$'s; and the
operator $\psid\sigmav\cdot\Ev\times\Ev\psi$, which is nonzero in
nonabelian theories. Of these, only three interactions occur in our
treatment:
 \begin{eqnarray}  \label{spinsplit}
 \delta\L_{\rm spin} &\equiv&
    f_1\, \frac{g}{M^3}\,\psid\{\Dv^2,\sigmav\cdot\Bv\}\psi  \nl
 &+& f_2\,\frac{ig}{M^4}\,\psid\{\Dv^2,\sigmav\cdot(\Dv\times\Ev -
     \Ev\times\Dv)\}\psi  \\
 &+& f_3\,\frac{i g^2}{M^3}\,\psid\sigmav\cdot\Ev\times\Ev\psi \, . \nonumber
 \end{eqnarray}

The power-counting rules indicate that spin splittings in quarkonium
systems should be smaller than splittings between radial and orbital
excitations. This is a familiar feature of the experimental data. The
$\psi^\prime$--$\psi$ and $\Upsilon^\prime$--$\Upsilon$ mass splittings are
both around 600~MeV. Our analysis indicates that spin splittings should
be smaller than this by a factor of $v^2$; that is, approximately
180~MeV for $\psi$'s and 60~MeV
for $\Upsilon$'s. In fact experiments show that the
$p$-state hyperfine splitting, $\chi(2^{++})$--$\chi(0^{++})$, is
141~MeV for the $\psi$~family, and 34~MeV for $\Upsilon$'s. The sizes of
these splittings are consistent with our expectations, and show clear
evidence of appropriate scaling with $v^2$. The $s$-state hyperfine
splitting between the $\psi$ and $\eta_c$ of about 117~MeV is also
consistent. These data reinforce our confidence in the power-counting rules,
and in the utility of a nonrelativistic framework for studying these mesons.

One final source of corrections comes from vacuum polarization of heavy
quarks. These modify the gluon action at order $\alpha_s(M)\,v^2$ and so can
probably be ignored at the 10\% level. Such corrections are essentially
relativistic in character and should not be analyzed using NRQCD. They are
easily computed using continuum perturbation theory.

\section{Relativistic Corrections}

\label{rel}

The corrections to the continuum NRQCD action required by relativity
were enumerated
in \secn{pcounting}. Our task now is to find values for the coupling
constants multiplying these operators such that the predictions of NRQCD
agree with those of ordinary QCD through order $v^2$. Since the coupling
constants have perturbative expansions, we will determine them by matching
perturbative results in the two theories. Here we work only to lowest order,
or tree level, in perturbation theory. Higher order corrections will be
discussed in \secn{alpha}.

\subsection{Kinetic Terms}

\label{kinsect}

The simplest correction follows immediately from the formula for the
relativistic energy of a noninteracting quark,
 \begin{equation}  \label{kinetic}
 \sqrt{\pv^2 + M^2} \approx M + \frac{\pv^2}{2M} - \frac{\pv^4}{8M^3}\, .
 \end{equation}
This implies a correction term, appropriately gauged, of the
form
 \begin{equation}
 \label{dlkin}
 \delta\L_K =  \frac{1}{8M^3}\psid\Dv^4\psi \, ,
 \end{equation}
which fixes $c_1 = 1/8$ in \eq{bilinear}.

\subsection{Electric Couplings}

\label{ecoupsect}

Most of the corrections from \secn{pcounting} involving the
chromoelectric field are linear in $g\Ev$. We compute these terms
by examining the order $g$ amplitude~$T_E$ for  scattering a
quark off a static electric field in QCD,
 \begin{equation}
 \label{Eamplitude}
 T_E(\pv,\qv) = \overline{u}(\qv) \gamma^0\, g\phi(\qv - \pv) \, u(\pv)\, ,
 \end{equation}
with $\phi$ the scalar potential.  We are interested in matching
this result at small $v$ to that obtained in NRQCD, and so
must expand it in powers
of $\qv/M$ and $\pv/M$. The Dirac spinor,
normalized nonrelativistically with $u^\dagger u=1$,
is
 \begin{equation}
 u(\pv) = \left(\frac{E_p+M}{2E_p}\right)^{\mbox{$\frac{1}{2}$}}
 \left[
 \begin{array}{c}
 \psi \\[.12in]
 {\displaystyle {\sigmav\cdot\pv\over E_p+M}}\,\psi
 \end{array}
 \right] \, ,
 \end{equation}
with $E_p\equiv\sqrt{\pv^2+M^2}$ and $\psi$ a two-component spinor.
Substituting this expression in \eq{Eamplitude} we obtain
 \begin{eqnarray}
 T_E(\pv,\qv) & = & \sqrt{\frac{(E_p + M)(E_q + M)}{4 E_p E_q}} \nl
        &\times&\psid\, \left[ 1 + \frac{\pv \cdot \qv + i \sigmav \cdot
        \qv \times \pv}{(E_q + M) (E_p + M)} \right] \, g \phi(\qv-\pv)\, \psi
             \\[.2in]
             & \equiv & S_E(\pv,\qv) + V_E(\pv,\qv) \, ,   \nonumber
 \end{eqnarray}
where $S_E$ denotes the scalar part of the amplitude, independent of $\sigmav$,
and $V_E$ is the spin-dependent term.
Expanding both $S_E$ and $V_E$ to second order in
$\pv/M$ and $\qv/M$,
 \begin{eqnarray}
 S_E(\pv,\qv) & = & \left( 1 - \frac{(\pv -\qv)^2}{8M^2} \right)
                     \psid \; g \phi(\qv-\pv) \;\psi \\
 V_E(\pv,\qv) & = & \left(\frac{i}{4M^2} - \frac{3 i}{32 M^4} (\pv^2 + \qv^2)
                     \right) \psid \;\sigmav \cdot \qv
                         \times \pv \; g \phi(\qv-\pv) \;\psi \, .\nonumber
 \end{eqnarray}

By computing the same process in NRQCD and comparing with the QCD result at low
momentum, it is possible to determine which operators must appear in the
NRQCD lagrangian and to fix their coefficients.  The first term in the
lowest order NRQCD action, \eq{leading}, predicts a scattering
amplitude $\psid\, g\phi(\qv-\pv)\, \psi$ .  This
corresponds to the first term in $S_E$.  The remaining terms in $S_E$ and
$V_E$ can only be included by adding the new interactions
 \begin{eqnarray}
 \label{dlele}
 \delta \L_E & = & \frac{g}{8 M^2} \psid
                       (\Dv \cdot\Ev - \Ev \cdot\Dv) \psi \nl
             & + & \frac{i g}{8 M^2} \psid \left( \sigmav \cdot \Dv \times \Ev
-
                   \sigmav \cdot \Ev \times \Dv \right) \psi \\
             & + & \frac{3 i g}{64 M^4} \psid \{ \Dv^2,\,
                   \sigmav \cdot \Dv \times \Ev -
                   \sigmav \cdot \Ev \times \Dv \} \psi \, .\nonumber
 \end{eqnarray}
to the NRQCD action.
The coefficients $c_2$ and $c_3$ in \eq{bilinear} are therefore
$1/8$, while $f_2$ in \eq{spinsplit} is $3/64$.

\subsection{Magnetic Couplings}

Similarly, we determine the operators linear in $\Bv$
by calculating the amplitude for the scattering of a quark off a static
vector potential $\Av({\bf x})$. The time independence of $\Av({\bf x})$
guarantees that there is no electric component.
It also enforces the conservation of the quark energy.

The amplitude is
 \beq
 T_B(\pv,\qv) = - \overline{u}(\qv) \gammav \cdot g \Av(\qv-\pv) u(\pv) \, ,
 \eeq
where $\pv^2 = \qv^2$.
Expanding in $\pv/M$ and employing energy conservation,
 \beqa
 T_B(\pv,\qv) & = & - \frac{g}{2 M} \left( 1 - \frac{\pv^2}{2 M^2} \right)  \nl
              & \times & \psid \left[ (\pv + \qv) \cdot \Av +
                      i \sigmav \cdot \Av \times (\pv - \qv) \right] \psi
\\[.2in]
              & \equiv & S_B(\pv,\qv) + V_B(\pv,\qv) \, . \nonumber
 \eeqa
The spin-independent term $S_B(\pv,\qv)$ arises in NRQCD from the terms
linear in $g\Av$ which appear in the kinetic energy part of the action,
Eqs.~(\ref{leading}) and (\ref{dlkin}).  The spin-dependent term
$V_B$ must be generated by the new interactions
 \beqa
 \label{dlmag}
 \delta \L_B & = & \frac{g}{2 M} \psid \sigmav \cdot \Bv \psi \nl
             & + & \frac{g}{8 M^3} \psid \{ \Dv^2, \,
                       \sigmav \cdot \Bv \} \psi \, ,
 \eeqa
so that $c_4 = 1/2$ and $f_1 = 1/8$ in \eq{bilinear} and \eq{spinsplit} .

\subsection{Bilinears in $F_{\mu\nu}$}

\label{bilinsect}

The interaction $\psid \,\sigmav\cdot\Ev\times\Ev \,\psi$,
which contributes to spin splittings, is the only bilinear in  $F_{\mu\nu}$
that contributes to the accuracy desired.
To fix the coefficient of this operator ($f_3$ in \eq{spinsplit}), we
calculate the amplitude for double scattering of a quark off an external
static electric field. In QCD, it is
\beqa
\label{teefull}
T_{EE}(\pv_1,\pv_2) &=& - i \int {d^3 q_1\over(2\pi)^3}
{d^3 q_2\over(2\pi)^3} \, (2\pi)^3 \delta^3(\qv_1+\qv_2+\pv_1-\pv_2)  \\
&\times& \overline{u}(\pv_2)\gamma^0\, g\phi(\qv_2)\,
  \frac{i}{(\rlap/p_1 + \rlap/q_1)-M+i\epsilon}\,
   g\phi(\qv_1)\,\gamma^0 u(\pv_1)\, . \nonumber
\eeqa
This same process must be computed in NRQCD, and compared to an expansion of
\eq{teefull} at small $\pv /M$.  This comparison is simplified by using
the identity
\beq
 \label{idprop}
 \frac{1}{\rlap/k - M} = \sum_s\left[\frac{u_s(\kv)\,\overline{u}_s(\kv)}
     { k^0 - E} + \frac{v_s(-\kv) \,\overline{v}_s(-\kv)}{k^0 + E} \right]\,
\eeq
for the fermion propagator, to break $T_{EE}$ into two terms before
expanding.
NRQCD, with the terms computed thus far, reproduces
the first term containing $u\,\overline{u} /(k^0 - E)$
to the required order in $v$, using vertices derived from the interactions of
\eq{dlele}.
These were designed to match the matrix element of \eq{Eamplitude},
$\overline{u}\gamma^0\,g\phi\,u$, and the first term of \eq{idprop}
contributes the same matrix elements to each of the vertices of the
QCD diagram.  The energy denominator
\beq
\frac{1}{k^0 - E} \sim \frac{1}{\tilde{k}^0 - \kv^2/2M} -
\frac{1}{\tilde{k}^0 - \kv^2/2M}
\left(\frac{\kv^4}{8 M^3}\right)\frac{1}{\tilde{k}^0 - \kv^2/2M} \, ,
\eeq
with the kinetic energy $\tilde{k}^0\equiv k^0-M$, is reproduced
in NRQCD by its nonrelativistic propagator and the kinetic
term correction of Section \ref{kinsect}.

The second part of $T_{EE}$, when expanded, introduces a term of an
entirely different character.  To lowest order in $v$, for low-momentum
external
gluons, the antiquark propagator
is just $(2M)^{-1}$, so this part behaves as a local
seagull interaction.  This contributes to \eq{teefull}
\beqa
&& - \int {d^3 q_1\over(2\pi)^3}
{d^3 q_2\over(2\pi)^3} \, (2\pi)^3 \delta^3(\qv_1+\qv_2+\pv_1-\pv_2)  \\
&\times& \frac{1}{8M^3}
\psid \, g\phi(\qv_2)\,
[\qv_1\cdot\qv_2 - i\sigmav\cdot\qv_1\times\qv_2]
   g\phi(\qv_1)\,\psi \, . \nonumber
\eeqa
In order to obtain the same result with NRQCD, we must include the
new interaction
\beq
 \label{dlbil}
 \delta\L_{EE} = -\frac{g^2}{8M^3}
       \psid\,[\Ev\cdot\Ev + i\sigmav\cdot\Ev\times\Ev]\psi\, .
\eeq
To the order to which we are working only the second term,
which contributes to spin splittings at order $v^4$, must be retained.
So $f_3 = -1/8$ \eq{spinsplit}, and we have determined all
relevant tree-level coefficients of the NRQCD
lagrangian.

\subsection{The NRQCD Continuum Lagrangian }

The lagrangian
for continuum NRQCD in Minkowski space is
 \beqa
 \label{lagrangian}
 \L_{NRQCD} & \equiv & i \psid D_t \psi + \psid \frac{\Dv^2}{2 M} \psi
\nl[.1in]
            & + & \! \delta \L_K + \delta \L_E + \delta \L_B + \delta \L_{EE}
\, ,
 \eeqa
where the relativistic corrections are defined in Eqs.~(\ref{dlkin}),
(\ref{dlele}), (\ref{dlmag}) and (\ref{dlbil}), respectively.
We have not included in the lagrangian a mass term $M\psid\psi$ because in a
nonrelativistic framework it only fixes the zero-point energy
and has no effect on mass splittings, wavefunctions, and so on.
As an independent check, we performed the systematic but lengthy
calculation of the lagrangian in \eq{lagrangian} using
the Foldy-Wouthuysen-Tani transformation \cite{foldy}.  This transformation
is not directly applicable beyond tree level.  However, perturbative
corrections to the various coefficients may be computed by comparing
amplitudes, as in Sections~\ref{ecoupsect} - \ref{bilinsect},
but including loop corrections.

Finally, we note that for lattice simulations we will need the
Euclidean action. This is obtained from the Minkowski theory defined by
\eq{lagrangian} by keeping track of how the three-dimensional vectors and
scalars are defined in terms of Lorentz-covariant quantities. Under
Wick rotation the zero component of contravariant vectors
rotates as the time coordinate,
 \beq
 x^0_{(M)} = - i x^0_{(E)} \, ,
 \eeq
while the zero component of covariant vectors
rotates as the time derivative,
 \beq
 \partial_0^{(M)} = + i \partial_0^{(E)} \, .
 \eeq
Spatial components of all four-vectors are unchanged.
Wick rotation changes the sign of the $(0,0)$ component of the metric tensor,
so that the usual Minkowski metric $\eta_{\mu \nu} \equiv {\rm diag}
(1, -1, -1, -1)$ rotates to the negative of the identity matrix. One then has
to
extract this minus sign from scalar products in order to work with the usual
positive definite euclidean metric.
Our definitions for the basic operators appearing in \eq{lagrangian} are
 \beqa
 \label{euclidean}
 D_0^{(M)} &=& + i D_0^{(E)} \nl
 \phi_{(M)}  &=& - i \phi_{(E)} \\
 E^i_{(M)} & = & - i E^i_{(E)}    \nonumber \, ,
 \eeqa
while other fields are unaffected.

\section{Lattice NRQCD}

\label{latt}

\subsection{Leading Order}

NRQCD is readily reformulated on a discrete space-time lattice. As usual,
quark fields~$\psi(x)$ are defined at the nodes of the lattice, while gauge
fields are represented by unitary matrices~$U_{x,\mu}$ defined on the
links joining neighboring nodes. Covariant derivatives are replaced by
forward, backward or centered differences,
 \begin{eqnarray}
 a\dplus\mu \psi(x) &\equiv& \U x\mu \,\psi(x+a\hat{\mu}) - \psi(x) \nl
 a\dminus\mu \psi(x) &\equiv& \psi(x) -
  \Udag{x-a\hat{\mu}}\mu \,\psi(x-a\hat{\mu}) \nl
 \dpm{} &\equiv& \mbox{$\frac{1}{2}$}(\dplus{} + \dminus{})\, ,
 \end{eqnarray}
depending upon the details of the interaction, while the laplacian becomes
 \begin{equation}
 \dsq \equiv \sum_i \dplus i\dminus i  = \sum_i \dminus i\dplus i.
 \end{equation}
The field $\F\mu\nu(x)$ is efficiently represented by
cloverleaf operators defined at the nodes of the lattice \cite{sheik} ,
 \begin{equation} \label{cloverleaf}
 g \Fc\mu\nu(x) = - \frac{1}{4a^2}
 \sum_{P(x,\mu\nu)} {\cal I}[U_{P(x,\mu\nu)}] \, ,
 \end{equation}
where the sum is over all plaquettes~$P$ in the $(\mu,\nu)$~plane
containing the site~$x$. $U_P$ is the product of link matrices on~$P$,
counterclockwise about~$\hat\mu\times\hat\nu$, as depicted in Fig. 1.
The operator $\cal I$ is defined as
 \begin{equation}
 {\cal I}[M] \equiv \frac{M-M^\dagger}{2i} -
 \frac{{\bf 1}}{3} \, {\rm Im}({\rm Tr} \, M) \, .
 \end{equation}
Covariant derivatives of $\F\mu\nu$ are represented by the differences
 \begin{eqnarray} \label{gaugederiv}
 a\dplus\rho \Fc\mu\nu(x) &\equiv& \U x\rho \, \Fc\mu\nu(x+a\hat \rho)
    \Udag x\rho - \Fc\mu\nu(x) \nl
 a\dminus\rho \Fc\mu\nu(x) &\equiv& \Fc\mu\nu(x) - \Udag {x-a\hat\rho}\rho
 \Fc\mu\nu(x-a\hat\rho) \U {x-a\hat\rho}\rho \, .
 \end{eqnarray}
All terms in the NRQCD action are built with these elements.

Focusing on the leading-order terms, we may define a series
of lattice actions for NRQCD,
 \begin{eqnarray}
 S^{(n)}_0 &=& a^3 \sum_x \psi^\dagger(x)\psi(x) \nl
 &-& a^3 \sum_x \psi^\dagger(x+a\hat t)
\left( 1 - \frac{aH_0}{2n}\right)^n\,\Udag x t \,
\left( 1 - \frac{aH_0}{2n}\right)^n\,\psi(x) \, ,
 \end{eqnarray}
where $n$ is a positive integer and $H_0$ is the kinetic energy operator
 \begin{equation}
 H_0 \equiv -\frac{\dsq}{2M}\, .
 \end{equation}
The quark Green function for such a theory satisfies a simple evolution
equation,
 \begin{equation} \label{evoleqn}
 G(\xv,t+a) \, =\, \left( 1 - \frac{aH_0}{2n}\right)^n \Udag x t \,
 \left( 1 - \frac{aH_0}{2n}\right)^n G(\xv,t) +
                    \delta_{\xv,0}\delta_{t+a,0}\, ,
 \end{equation}
where $G(\xv,t)$ vanishes for $t<0$.
This evolution equation differs from others that have been suggested. It is
symmetric with respect to time reversal, unlike
that of \cite{leptactwo}.  Also, as we will see, it leads to a
wavefunction renormalization much smaller than that of \cite{davtac}.

The parameter~$n$ was introduced to prevent
instabilities at large momenta due to the kinetic energy operator
\cite{leptactwo}. The Green function defined by \eq{evoleqn} blows up if
${\rm max}(a H_0/2n) \ge 2$, and
so $n$~must be large enough to avoid this. Neglecting gluons, this
implies that $n>3/(2Ma)$ is required for stability; gluonic effects reduce the
3/2 to something closer to 1.15 for $\beta\approx 6$ \cite{davtac}.
Thus, at $\beta = 6$, $n=1$ suffices for $b$ quarks, while $n=2$ is needed for
$c$
quarks. Larger $n$'s may be required as $\beta$ is increased and $a$
decreases.

\subsection{Lattice Spacing Errors}

The finite spacing of the lattice introduces new systematic errors into NRQCD.
Errors associated with the temporal spacing~$a_t$ are typically of
order $a_t\langle{K}\rangle$, while those due to
the spatial lattice are of order $a_x^2\langle \pv^2 \rangle$.
The two tend to be roughly comparable since usually $a_x =
a_t \approx 1/M$. Both must be removed if we are to achieve high
precision.

Errors due to the finite lattice spacing  are removed by adding new
interactions to the lagrangian, just as $\O(v^2)$~errors were removed. As in
that case, we could compute the necessary corrections by matching
perturbative scattering amplitudes in lattice NRQCD with those in continuum
QCD. However, at tree level, it is simpler to correct the individual
components from which the lattice action is built ($\dsq$, $\Ev$,
$\Bv$\ldots) so that they more accurately reproduce the effects of their
continuum analogues. This is the approach we will follow here.

To compare the effects of lattice operators with those of continuum operators
we need some way to relate fields on the lattice to those in the
continuum. For the quarks, we simply identify the lattice field at a site
with the continuum field at the same space-time point. For the gluons, gauge
invariance requires that the link operator be related to the continuum gauge
field through the path-ordered exponential of a line integral,
 \beq
 \label{link}
 U_{x,\mu} \equiv P \, \exp \left[\, - i g \int_{x}^{x+a \hat{\mu}}
              A \cdot dy \, \right] \, ,
 \eeq
where the scalar product is taken with the positive euclidean metric.
The simplest choice of path for the integral is a straight line joining
$x$ to $x+a \hat{\mu}$; other choices lead to the same final results but
give more complicated $\O(a)$ corrections. With this mapping from lattice
variables to continuum variables, we can now correct each of the lattice
operators.

\subsubsection{Spatial Derivatives}

Given relation~(\ref{link}), the action of the lattice difference operators
on continuum fields is specified by
 \beqa
\label{exactdef}
 a\dplus i \equiv& \exp(aD_i)-1 &= aD_i + \frac{a^2}{2} D_i^2 +\ldots\nl
 a\dminus i \equiv& 1 - \exp(-aD_i) &= aD_i - \frac{a^2}{2} D_i^2 +\ldots\; .
 \eeqa
These are just gauge-covariant extensions of the obvious $g=0$ relations.
By combining these expansions we obtain an improved difference
operator that reproduces the effects of $D_i$ through order $a^4$:
 \beq \label{imprD}
 \tilde{\Delta}^{(\pm)}_i
\equiv \dpm i - \frac{a^2}{6}\dplus i \dpm i \dminus i.
 \eeq
Similarly we may define an improved lattice laplacian that is also accurate
to order $a^4$:
 \beq \label{axcor}
 {\bf\tilde\Delta}^{(2)} = \dsq - \frac{a^2}{12}\sum_i\left[\dplus i
 \dminus i \right]^2 .
 \eeq

\subsubsection{Temporal Derivative}

Temporal and spatial derivatives
enter NRQCD differently because the theory is nonrelativistic.  One
consequence is that quark propagation is governed by a Schr\"odinger equation
with a single time derivative. The calculation of quark Green functions
is therefore an initial value problem, which is much less costly to
solve numerically than the boundary value problem dictated by the Dirac
equation. If we improve the time derivative operator as we did
for spatial derivatives, this simplification is lost,
as we would need to  introduce higher order time derivatives.

To find an alternative, we examine the evolution equation for the
quark Green function, \eq{evoleqn}. Neglecting the gauge field for the
moment, this equation may be written (for $t>0$)
 \beqa
 G(\xv,t+a) &=& \left(1-\frac{aH_0}{2n}\right)^{2n} G(\xv,t) \nl [.1in]
 &=& \e^{-aH_\eff} G(\xv,t) \, ,
 \eeqa
where the effective hamiltonian
 \beqa
 H_\eff &\equiv& - \frac{2n}{a}\,\ln\left(1-\frac{aH_0}{2n}\right) \nl
 &=& H_0 + \frac{a}{4n}\,H_0^2 +\cdots\; .
 \eeqa
The order~$a$ term in the effective hamiltonian cancels if we replace $H_0$
by
 \beq \label{atcor}
 \tilde{H}_0 \equiv H_0 - \frac{a}{4n}H_0^2 \, .
 \eeq
Making this replacement in the full evolution equation (\eq{evoleqn}) removes
the leading error due to our lattice approximation of the temporal
derivative.\footnote{Only the kinetic part of the hamiltonian needs
fixing here. The gauge-potential part is automatically exponentiated since
the gauge fields enter through the link variables~$\U x\mu$.} The resulting
theory is equivalent to what would have been obtained by improving the
temporal derivative along the lines presented in the previous section; the two
theories are related by a redefinition of the quark field, and so give the
same results for energy levels, on-shell scattering amplitudes and other
physical quantities.

\subsubsection{Electric and Magnetic Fields}

The lattice cloverleaf field $\Fc\mu\nu(x)$ (\eq{cloverleaf}) is equal to the
continuum field~$\F\mu\nu(x)$ up to corrections of order~$a^2$. Our
power-counting analysis (\secn{pcounting}) implies that only corrections
linear in $\F\mu\nu$ are important. Such corrections are the same in an
abelian theory as they are in QCD, so we simplify the analysis
by focusing on the abelian case.

In the abelian theory, where the operator $\cal I$ just takes the imaginary
part, the cloverleaf field is
 \beqa
 \label{stokes}
 a^2 g F^{(c)}_{\mu \nu} (x) & = & - \; {\rm Im} \left( 1 - \frac{i g}{4}
           \oint_{(2 \times 2)} A \cdot dy + \cdots \right)  \nl
 & = & \frac{g}{4} \int_{(2 \times 2)} (\partial_\mu A_\nu -
        \partial_\nu A_\mu)
          \; dy^\mu \wedge dy^\nu + \O (a^6) \\
 & = & a^2 g F_{\mu \nu} (x) + \frac{a^4}{6}
        (\partial^2_\mu + \partial^2_\nu) g F_{\mu \nu} (x) + \O (a^6)
        \nonumber \, ,
 \eeqa
where the surface integral is over the $(2 \times 2)$ plaquette in the
$(\mu,\nu)$ plane centered at $x$. The $a^4$ term comes from the
second-order term in the  Taylor expansion
of $F_{\mu \nu}$ around point $x$.
The nonabelian generalization of this result is obtained by
replacing derivatives with covariant derivatives.

By subtracting the lattice version of the $\O(a^2)$ error,
 \beq
\label{corrclov}
 g \tilde{F}^{(c)}_{\mu \nu} (x) = g F^{(c)}_{\mu \nu} (x) -
      \frac{a^2}{6} \left[ \Delta^{(+)}_\mu \Delta^{(-)}_\mu +
       \Delta^{(+)}_\nu \Delta^{(-)}_\nu \right] g F^{(c)}_{\mu \nu} (x) \, ,
 \eeq
we have a definition accurate to $\O(a^4)$. Expanding the lattice
derivatives as in \eq{gaugederiv}, we obtain finally
 \beqa
\label{newcorrclov}
 g \tilde{F}^{(c)}_{\mu \nu} (x) & = & \frac{5}{3} \, g F^{(c)}_{\mu \nu} (x) -
     \frac{1}{6} \left[ U_{x, \mu} g F^{(c)}_{\mu \nu} (x + a \hat{\mu})
       U^{\dagger}_{x, \mu}  \right. \nl
         &  & \left. + \; U^{\dagger}_{x - a \hat{\mu}, \mu}
         g F^{(c)}_{\mu \nu} (x - a \hat{\mu}) U_{x - a \hat{\mu}, \mu} \, - \,
          \left( \mu \Leftrightarrow \nu \right) \right] \, .
 \label{imprF}
 \eeqa
We have verified that $\tilde{F}^{(c)}_{\mu \nu}$ reproduces the continuum
$F_{\mu \nu}$ to $\O(a^4)$ directly in the nonabelian theory using
techniques described in the Appendix.

\subsubsection{The Gauge Field Action}

The gluon action has lattice spacing errors as large as those of
the nonrelativistic quark action. A great deal of sophisticated work
has been done on improving the gauge field action,
dealing mainly with one-loop contributions \cite{improve}.
For us, however, a
straightforward improvement of the classical action is all that is required.
In particular, corrections involving nonplanar loops
are not needed.  This greatly simplifies our analysis.

Power counting indicates that the nonabelian and abelian
corrections have the same form.  By
Stokes'  theorem, a single abelian plaquette centered at point $x$
is related to the continuum field tensor by
 \beq
 \label{oneplaquette}
 U_P (x) = \exp \left[ - i g a^2 F_{\mu \nu} (x) - i g \frac{a^4}{24}
    (\partial^2_\mu + \partial^2_\nu) F_{\mu \nu} (x) + \O (a^6) \right] \, ,
 \eeq
and the usual lattice lagrangian is
 \beq
 \label{reoneplaq}
 \frac{1}{2 g^2 a^4} {\rm Re} \left( U_P (x) - 1 \right) =
   - \frac{1}{4} F_{\mu \nu}^2 (x) - \frac{a^2}{48} F_{\mu \nu} (x)
        (\partial^2_\mu + \partial^2_\nu) F_{\mu \nu} (x) + \O (a^4) \, .
 \eeq
The nonabelian generalization is straightforward; the key result is that
the corrections are of order $a^2$.

A simple scaling argument allows us to correct the nonabelian action.
The usual lagrangian involves a sum over all four $(1\times 1)$ plaquettes at
a site~$x$,
 \beq
 \label{lgdef}
 \L_G (x) = \frac{1}{4 g^2 a^4} \sum_{P(1\times 1)}
                  {\rm Tr} \left[ \R \left( U_{P(1\times 1)} \right) \right],
 \eeq
 where
 \beq
 \R(M) \equiv \frac{M + M^{\dagger}}{2} - 1 \, .
 \eeq
{}From the discussion above,
 \beq
 \label{scale}
 \L_G = \L_G^{\rm cont} + a^2\, \delta \L_G + \O(a^4) \, ,
 \eeq
where $\L_G^{\rm cont}$ is the continuum lagrangian. An acceptable lagrangian
may also be constructed from $(2\times 2)$ plaquettes, and its $a$-dependence
follows from Eqs.~(\ref{lgdef}) and~(\ref{scale}), with the replacement $a\to
2a$:
 \beqa
 \L_G^{(2\times 2)} (x) &=& \frac{1}{64 g^2 a^4} \sum_{P(2\times 2)}
           {\rm Tr} \left[ \R \left( U_{P(2\times 2)} \right) \right]\nl
 &=& \L_G^{\rm cont} + 4 a^2\, \delta \L_G + \O(a^4)\, .
 \eeqa
By combining these two lagrangians, we obtain a gluon lagrangian
that is accurate to order~$a^4$,
 \beqa
 \tilde\L_G &=& \frac{1}{3} \left( 4 \L_G - \L_G^{(2\times 2)}
 \right) \nl
  & = & - \mbox{$\frac{1}{2}$} {\rm Tr} \left[ F^2_{\mu \nu} \right] +
  \O(a^4)\, .
 \eeqa
Again, this result may be confirmed
directly in the nonabelian theory by means of the techniques described
in the Appendix.

\subsection{A Corrected Evolution Equation}

\label{correvoleqn}

The leading-order evolution equation, \eq{evoleqn}, for the quark Green
function is readily generalized to include relativistic and finite~$a$
corrections.  We choose the full evolution equation to be
 \beqa
 \label{fullev}
 \lefteqn{G(\xv,t+a) =} \nl
 & & \left( 1 - \frac{a\tilde H_0}{2n}\right)^n
 \left(1-\frac{a\,\dH}{2}\right) \Udag x t \,\left(1-\frac{a\,\dH}{2}\right)
 \left( 1 - \frac{a\tilde H_0}{2n}\right)^n G(\xv,t)
 \eeqa
for $t>0$. Here we have corrected the leading kinetic energy operator for
finite lattice spacing errors as in Eqs.~(\ref{axcor}) and~(\ref{atcor}) so
that
 \beq
 \tilde H_0 \equiv - \frac{{\bf\tilde\Delta}^{(2)}}{2M}
  - \frac{a}{4n} \frac{(\dsq)^2}{4M^2}\, .
 \eeq
Further, a straightforward transcription of the
relativistic corrections into lattice
variables gives, for the spin-independent corrections,
 \beqa
 \dH_{K,v^2} &\equiv& -\frac{(\dsq)^2}{8 M^3} \\
 \dH_{v^2} &\equiv& \frac{ig}{8 M^2}\, (\Deltavpm\cdot\Ev
 - \Ev\cdot\Deltavpm)\, ,
 \eeqa
and, for the spin-dependent terms,
 \beqa
 \dH_{\rm spin} &\equiv&  - \frac{g}{8 M^2}\,\sigmav\cdot(
 \Deltavtpm\times\tilde\Ev - \tilde\Ev\times\Deltavtpm)  \nl
 &-& \frac{g}{2 M}\, \sigmav\cdot\tilde\Bv \\
 \dH_{{\rm spin},v^2} &\equiv&
     -\frac{g}{8 M^3}\,\{\dsq,\sigmav\cdot\Bv\}  \nl
 &-& \frac{3g}{64 M^4}\,\{\dsq,\sigmav\cdot(\Deltavpm\times\Ev -
     \Ev\times\Deltavpm)\}  \nl
 &-& \frac{i g^2}{8 M^3}\,\sigmav\cdot\Ev\times\Ev \, .
\eeqa
Here  $\Ev$ and $\Bv$ are defined in terms of the cloverleaf
field, Eqs.~(\ref{cloverleaf}) and~(\ref{imprF}),
 \beqa
\label{EBvsF}
 E^i (x) & = & F_{0i}^{(c)} (x) \\
 B^i (x) & = & \mbox{$\frac{1}{2}$} \epsilon_{i j k } F_{jk}^{(c)} (x) \, .
 \eeqa
The difference operators in these terms act on all the fields, $\Ev$,
$\Bv$, and $G(\xv,t)$, to their right. However, the $\Dv\cdot\Ev-\Ev\cdot\Dv$
operator may be simplified since the net effect is to differentiate only the
$\Ev$~field. Using \eq{gaugederiv}, it can be written as
 \beq
\Deltavpm\cdot\Ev
= \frac{1}{2 a} \, \sum_{i}
\left[\,  U_{x,i} E^i (x+a \hat{\imath})  U^\dagger_{x,i}
-  U^\dagger_{x-a \hat{\imath},i} E^i (x-a \hat{\imath})
      U_{x-a \hat{\imath},i} \,\right]  \, .
 \eeq
Finite lattice spacing corrections are included in $\dH_{\rm spin}$ by using
$\tilde\Ev$ and $\tilde\Bv$ of \eq{imprF} and $\Deltavtpm$ of \eq{imprD} in
place of $\Ev$, $\Bv$ and $\Deltavpm$. These are unnecessary in the
other operators, which are already sufficiently accurate.

Certain correction terms may be unimportant in some calculations. The
$\O(v^2)$ corrections for spin-independent quantities are all included in
$\dH_{K,v^2}$, $\dH_{v^2}$, and $\dH_{{\rm spin}}$; the term $\dH_{{\rm
spin},v^2}$ enters only at order $v^4$. Spin splittings, however, are
themselves $\O(v^2)$, and so a comparably accurate measurement of such a
splitting would require all of the corrections. For simulations of
light-heavy mesons, such as $B$'s or $D$'s, probably only the magnetic part
of $\dH_{\rm spin}$ is important. All other terms are suppressed by
at least an additional power of the heavy quark velocity.

For reasons of efficiency we have separated the relativistic
corrections~$\dH$ from the leading order kinetic
energy~$\tilde H_0$ in \eq{fullev}. This
is possible because terms proportional to $\Ev$ or $\Bv$ do not cause
instabilities at high momenta; it is only the kinetic terms that cause
problems. The relativistic correction to the kinetic
energy~$\dH_{K,v^2}$ is a special case.
It probably should not be included in the
evolution equation at all, but rather its contribution computed in
first-order perturbation theory. The problem with this term is that it is
negative and large for large momenta, eventually overwhelming the leading
kinetic term~$\tilde H_0$ and destabilizing the theory. This problem is not
particular to the lattice version of NRQCD: it is a complication due to
the nonrenormalizability of the theory. In fact, the continuum hamiltonian
 \beq
 H = \frac{(\pv+g\Av)^2}{2M} - \frac{(\pv+g\Av)^4}{8M^3}
 \eeq
is not bounded below and will lead to problems if $\pv$ is allowed to become
too large. This problem does not arise in first-order perturbation theory,
so that a perturbative analysis of this correction is possible. Note also that
in
Coulomb gauge the correction can be approximated by $-\pv^4/8M^3$. Since
this is independent of the gauge field, its perturbative contribution to
mass splittings can be reliably determined given only the quark wavefunctions,
and these are easily computed in a simulation. Alternatively, one could
spread $\dsq$ for this term over, say, five lattice spacings in each
direction instead of three. This would reduce the largest effective momentum
by a factor of two, thereby reducing the high-momentum value of
$\dH_{K,v^2}$ by a factor of 16 while leaving the low-momentum behavior
largely unchanged.

Many of the same problems could arise from the finite lattice spacing
corrections in $\tilde H_0$. However it seems that these do not cause
problems for relevant values of~$Ma$. Indeed these terms seem likely to
reduce the maximum value of $H_0$.

A more radical way of dealing with instabilities
at high momenta might be to choose Coulomb gauge, where the vector potential
is small, and then replace
 \beq
 \left(1-\frac{aH_0}{2n}\right)^n \to
\exp\left[- \frac{a}{2}\,\left(\sqrt{{\pv}^{2} + M^2} - M\right)\right]
 \eeq
in the evolution equation. Here ${\pv}^{2}$ is the quark momentum, which is
easily introduced into the simulation through the use of Fast Fourier
Transforms. In this way the exact relativistic energy is incorporated into
the simulation in a way that avoids instabilities. Unfortunately this
approach greatly complicates the perturbative analysis of the theory, not
least because it breaks gauge invariance. Nevertheless such problems are
probably not insurmountable, and the approach may someday be useful.

Finally, note that four-fermion interactions as in \eq{contact}
may be incorporated naturally into the evolution equation
by replacing them with new ones which are quadratic in fermion
fields.   This is accomplished by introducing an auxiliary field $\eta$,
a complex matrix field which is $2\times 2$
in spin and $3\times 3$ in color indices.
For example, if $\eta$ appears in the lagrangian as
\beq \L_\eta = a\,\psid \eta \psi + b\, \chid \etad \chi +
         c\, {\rm Tr}\, \etad \eta
\eeq
without a kinetic term, this is equivalent to the interaction
\beq
\L_{\rm contact}  = -\frac{ab}{c}\,\psid\chi\chid\psi
\eeq
for the fermions.
This may be seen by performing explicitly the
gaussian path integral over $\eta$, or from the equations of motion
\beq
\eta = -\frac{b}{c}\,\chi\chid \qquad
\etad = -\frac{a}{c}\,\psi\psid \, .
\eeq
In a simulation this amounts to generating
and summing over a random gaussian distribution
for $\eta$ at each $\xv$ and $t$, normalized such that
\beq
<\eta_{ij} \eta_{kl} > = 0 \qquad < \eta_{ij} \eta^*_{kl} > =
\delta_{ik} \delta_{jl} \, ,
\eeq
and propagating the quark in this field according to the
interactions in $\L_\eta$.
\def\mf{{\rm mf}}

\section{Radiative Corrections}

\label{alpha}

In previous sections we derived tree-level values for the
couplings of NRQCD.  The couplings are modified by radiative
corrections that are dominated by momenta of order $\pi/a$ or larger.
Since $\pi/a$ is typically several GeV in simulations, we should
be able to compute these radiative corrections using
weak-coupling perturbation theory.  However, lattice
perturbation theory is notorious for its poor convergence,
which is much worse
than continuum QCD at comparable momenta.  The large corrections,
which come from tadpole and related diagrams
and spoil lattice perturbation theory,
apparently result from the nonlinear connection between lattice
link variables and the continuum gauge potential.
In the worst cases such
tadpole corrections are almost as large as tree-level
contributions, and some sort of nonperturbative
treatment becomes necessary.  Fortunately, tadpole contributions  are well
described by a mean-field approximation \cite{lepmac}.
Such an approach allows us to
compute most of a radiative correction nonperturbatively in
terms of quantities that are easily measured in simulations.
In this section we first outline a mean-field analysis for
lowest-order NRQCD, giving  estimates for all
renormalization effects in that theory.  We also compare the
mean-field theory results with exact results to first order in  perturbation
theory.
Next we discuss a simple modification of lattice NRQCD that
effectively removes all tadpole contributions.  Finally we
comment briefly on the significance of nonperturbative
contributions to the couplings.

\subsection{Mean Field Theory and Lowest-Order NRQCD}

A simple way of tracing the effects of tadpoles is to replace
the link operators $U_{x,\mu}$ in the NRQCD propagator by a
number~$u_0$ representing the vacuum expectation value of~$U_{x,\mu}$. One
gauge-invariant definition of~$u_0$ is in terms of the plaquette operator
 \beq
 u_0 = \bra{0} \, \mbox{$\frac{1}{3}$}
    {\rm Tr}\, U_{\rm plaq} \, \ket{0}^{1/4}\, ,
\eeq
which becomes perturbatively
\beq
 u_0 = 1-0.083 g^2 - \cdots\; . \label{upth}
 \eeq
Other definitions are possible, but these give similar results and will not
be considered here. With this definition, $u_0$ is easily measured in a
simulation: for example, $u_0=1-0.122$ at $\beta=6$. The measured~$u_0$
includes tadpoles to all orders of perturbation theory, and perhaps also some
nonperturbative effects.

Having a mean value for $U_{x,\mu}$, we may proceed with a
mean-field analysis of the NRQCD propagator.  In this approximation, the
kinetic-energy operator is
\beq
H_0 \approx - \frac{1}{2 M a^2} \sum_i
\left[ u_0 e^{a \partial_i} + u_0 e^{- a \partial_i} - 2 \right] \, ,
\eeq
so that
\beq
H_0 \approx h_0 + u_0 \frac{\pv^2}{2M}
\eeq
for a quark with a low momentum $\pv$, where
\beq
h_0 = 3 \left(1-u_0\right)/Ma^2\, .
\eeq
Substituing this into \eq{evoleqn} for the Green function and replacing
the link variables by $u_0$'s, we obtain finally a mean field approximation
to the quark propagator
\beq
G^\mf(\pv,t) \approx
 u_0^{t/a} \left( 1 - \frac{a h_0}{2n} -
                  \frac{a u_0}{2n}\frac{\pv^2}{2M} \right)^{2nt/a} \, .
\eeq
In the low-momentum limit, we expect the continuum propagator
to have the form
\beq
G^{\rm cont}(\pv,t)
                = Z_\psi^\mf \exp \left[ - t \left( E_0^\mf +
                       \frac{\pv^2}{2 Z_M^\mf M} \right) \right]\, .
\eeq
Matching these, we obtain
\beq
E_0^\mf = -a^{-1} \ln \left[ u_0 \left(1-ah_0/2n\right)^{2n}\right]
\eeq
for the zero-point energy induced by tadpoles,
\beq
Z_M^\mf = u_0^{-1}\,\left(1 - a h_0/2n \right)
\eeq
for the mass renormalization, and
 \beq
 Z_\psi^\mf =  1
 \eeq
for the quark field renormalization. At $\beta=6$ with $Ma=2$ (roughly the
$b$ quark mass) and $n=1$, these renormalization parameters are
 \beqa
 E_0^\mf &\approx& 0.9~{\rm GeV} \nl
 Z_M^\mf &\approx&  1.04 \nl
 Z_\psi^\mf &=& 1.
 \eeqa
None of the renormalizations is particularly large
for leading-order NRQCD, at least for $\beta$ near 6.  This is despite the
theory's nonrenormalizability and consequent power-law
divergent renormalizations: for example, $ah_0\propto 1/a$. These
divergences cause problems as the lattice spacing~$a$ is reduced, but
there will be little need to reduce~$a$ once our
finite-$a$ corrections have been incorporated.

Renormalization parameters like $E_0$, $Z_M$, and $Z_\psi$ are needed
to interpret simulation results.  For example, the total mass  of a meson
with NRQCD energy $E_n$ is
\beq \label{qmass}
 M_n = 2 \left(Z_M M - E_0 \right) + E_n \, .
\eeq
Knowing $Z_M$ and $E_0$ from mean-field theory, perturbation
theory, or both, and $E_n$ from simulations, we can use this
expression to tune the NRQCD bare quark mass.
The wave function renormalization is important when designing
quark operators to describe such things as radiative
transitions or decays through quark annihilation.  For example,
the continuum operator $\chi^\dagger \psi$, describing
quark-antiquark annihilation, is well modeled by the lattice
operator $\chi^\dagger \psi/{Z}_{\psi}$.

We can calibrate the reliability of mean-field theory by comparing
mean-field results with exact calculations to first order in perturbation
theory. Such calculations have been performed by
Davies and Thacker \cite{davtac}, who have analyzed the NRQCD propagator
through one-loop order.  They defined the propagator by means of
the evolution equation
 \beqa
 \label{dandt}
 G(\xv,t+a) & = & \Udag x t \left( 1 - \frac{a\,H_0}{n} \right)^n\,G(\xv,t) +
 \Udag x t \,\delta_{x,0}\,\delta_{t,0} \nl
 G(\xv,t) & = & 0  \qquad \qquad  ( t \leq 0 ) \, ,
 \eeqa
which differs slightly
from our equation. The energy shift and mass renormalization
for this equation are identical to ours if $n$ is replaced by $n/2$ in our
equations. The wavefunction renormalization however is significantly
different: in mean-field theory, one finds $Z_\psi^\mf = (1-ah_0/n)^{-n}$
for the Davies-Thacker equation.
In Table 2 we list the ${\cal O}(g^2)$
coefficients for $E_0$, $Z_M$, and $Z_\psi$. We give both the exact
result, as computed by Davies and Thacker using perturbation theory,
and the mean-field estimate, obtained by replacing $u_0$ with its
perturbative expansion, \eq{upth}, in mean-field expressions for the
renormalization parameters. Since we expect nontadpole
contributions of one or two times
$\alpha_s/\pi$, or $0.03g^2$ to $0.05g^2$,
the agreement is excellent. A nontadpole contribution of this size is only
5--10\% of the tree-level contribution at $\beta=6$. Mean-field theory does
seem
to account well for radiative	corrections when they are large.

When perturbative results are
known, they can be used to correct the mean-field prediction.
For example, the zero-point energy $E_0$ for our propagator
at $a M = 1.5$ can be written as
\beq
a E_0 = - \ln \left[u_0 \left( 1 - a h_0/2 \right)^2 \right]
+ 0.07 \, \frac{g^2}{u_0^4}\, .
 \eeq
The correction $0.07g^2$ is the difference between the
perturbative and mean-field result in Table 2.  The factor
$1/u_{0}^{4}$ in the correction accounts for tadpole induced
renormalization of the bare coupling $g^2 = 6/\beta$,
also suggested by mean-field theory \cite{lepmac}.
With this expression, all that is needed
to complete the calculation is a measurement of the expectation value of the
trace  of the plaquette to determine $u_0$.

\subsection{Removing Tadpole Contributions}

Our mean-field analysis of lowest-order NRQCD is readily
extended to include the various corrections due to relativity
and finite lattice spacing.  There is however a simpler way
of dealing with the tadpoles. By dividing every link matrix by
$u_0$ before computing quark propagators, all tadpole contributions are
automatically removed from the simulation.  Using $U_{x,\mu}/u_0$ in place of
$U_{x,\mu}$ everywhere, the  tree-level couplings we have computed should be
accurate to within corrections of order $\alpha_s/\pi$;
that is, to 5--10\% at $\beta=6.0$.

The mean-field corrections introduced by our simple procedure are not always
small, particularly for operators that involve the cloverleaf definitions of
electric or magnetic fields.  These operators are modified by a
factor $u_0^{-4}$: for example, $\Bv \rightarrow \Bv/u_0^4$.  At
$\beta=6.0$, $u_0^{-4}$ is $1.7$, almost doubling the $\Ev$
and $\Bv$ fields.  Leaving out such factors leads to dramatic
underestimates of quantities, such as spin splittings, that involve
one or more powers of $\Ev$ or $\Bv$.

\subsection{Nonperturbative Effects}

As is true of all perturbative calculations, our analysis of
NRQCD couplings omits nonperturbative contributions.
A priori, we know little about the
nature of such contributions.  The one thing we may assert with
some confidence is that nonperturbative contributions are
smaller than perturbative contributions for momenta $\pi/a$
larger than a couple of GeV; that is, for $\beta > 5.7$.  This is the
fundamental assumption underlying all applications of
perturbative QCD, whether in the continuum or on the lattice.
As we have argued,  perturbative corrections to our
couplings are probably less than 10--15\% for reasonable $\beta$'s, and so
it seems likely that nonperturbative
effects are no larger than a few percent.  Since most physical
results are not extremely sensitive to the values of the couplings,
the nonperturbative effects are probably safely neglected.

Our mean-field analysis supports this view
by providing a toy model for nonperturbative effects.
The plaquette operator plays a key role in
this analysis since it determines the mean field $u_0$.  The
plaquette is one of the few operators whose nonperturbative
behavior is somewhat understood.  The expectation value of a
large Wilson loop with area $A$ contains a factor $\exp(- \sigma A)$
due to nonperturbative confinement.  Here $\sigma$ is about
$0.18~{\rm{}GeV}^2$, based upon phenomenologically motivated
quark-antiquark potentials.  There is empirical evidence that
this behavior persists for small loops, and even for the plaquette, whose
expectation value seems well described by
perturbation theory times a factor of $(1 - \sigma a^2)$, at
least for $\beta$ near $6.0$ \cite{lepmac}.  At $\beta=6.0$, $ \sigma a^2$ is
about $0.04$, making the nonperturbative part of the plaquette
about $10\%$ of the size of the ${\cal O}(g^2)$ part.  This is perhaps
some indication of the relative sizes of perturbative and nonperturbative
effects.

To continue with our toy model, we assume that the mean-field
parameter $u_0$ inherits a nonperturbative contribution
$-{\sigma a^2}/4$ from the plaquette:
\beq
u_0^{\rm toy} = u_0^{\rm pert} - \sigma a^2/4 \, .
\eeq
This amounts to less than $2\% $ of $u_{0}^{\rm toy}$ at
$\beta=6.0$,
and less at higher $\beta$'s, so it has a negligible effect
on renormalization constants like $Z_M$ and $Z_\psi$.
One place where one might worry about
nonperturbative corrections is in the very divergent zero-point
energy,
\beqa \label{nonpertE}
E_0 &\approx & \left(1 - u_0 \right) \left( 1/a + 3/Ma^2 \right) \nl
    &\rightarrow & E_0^{\rm pert} + \sigma a/4 + 3 \sigma/4M \, .
\eeqa
Here there is a contribution that does not vanish with $a$; factors
of $1/a$ generated by perturbative power-law divergent loops cancel
the factors of $a$ in the nonperturbative term.  However this
contribution amounts to only about 30~MeV for b quarks,
much less than the perturbative contribution of about 900~MeV at $\beta=6$.

Nonperturbative terms such as those in \eq{nonpertE} for $E_0$ mean
that the perturbative relation, \eq{qmass}, between quark mass
and meson mass cannot be exact for any lattice spacing.  But
for $\beta=6.0$, the uncertainty that results is only a few
percent of the $b$ quark mass.  Of course, ours is only a toy
model for nonperturbative effects.  We do not really know that
the nonperturbative part of $u_0$ is suppressed by $a^2$ rather
than, say, $a^4$ or $a^{1/2}$.  But we do know that it is
suppressed relative to the ${\cal O}(g^2)$ contributions.  Thus we can
use perturbative results to bound nonperturbative
contributions; as long as perturbative corrections are not individually
large, nonperturbative corrections cannot be very important.
In some cases, the use of the measured value of $u_0$,
rather than the calculated one, may account for some of the
nonperturbative physics. This is most likely the case, for example,
when renormalizing cloverleaf $\Ev$'s and $\Bv$'s with the
plaquette average.

Finally it is worth noting that the nonperturbative contribution
to the plaquette expectation value is almost comparable in magnitude
to the ${\cal O}(g^4)$ perturbative contribution at $\beta=6.0$.
Thus there seems little point in computing much beyond
first or second order in $g^2$.  Precision beyond a
few percent will probably require nonperturbative tuning of
couplings.

\section{Conclusions}

\label{concl}

Nonrelativistic QCD provides one of the most efficient frameworks for
simulating heavy quarks.  While an approximation to QCD, it may be
systematically improved; this paper provides initial steps
toward that goal.  We developed general
power-counting rules which allowed us to assess the relative importance of
various corrections, and which we can use to fine tune the theory
for specific applications. We computed all of the leading order
corrections required by relativity, both for spin-independent and
spin-dependent interactions. The theory was then adapted to the lattice,
including all leading-order corrections from finite lattice spacing.
Finally, we presented a simple mean-field procedure that
automatically incorporates the largest radiative corrections to the NRQCD
coupling constants. Our mean-field analysis also allowed us to estimate the
importance of nonperturbative contributions to the couplings.

In NRQCD, heavy-quark propagators are determined by a simple
evolution equation, avoiding the need for costly matrix inversion.
Our fully corrected evolution equation was presented in
\secn{correvoleqn}. Using this equation, together with the mean-field
improvement procedure described in \secn{alpha}, should give results for
$\psi$ and $\Upsilon$ mesons that are accurate to perhaps 10\% or better,
depending upon the measurement. The largest remaining errors
are probably due to uncalculated $\O(\alpha_s)$ corrections to the
NRQCD couplings beyond the mean-field contributions, and to light-quark
vacuum polarization. These errors can be removed in the near future. The
coupling constant corrections are computed using ordinary one-loop
weak-coupling perturbation theory; work has already begun on these.  The
dominant contribution from  light-quark vacuum polarization is probably
insensitive to light quark masses less than
100--200~MeV, and the extrapolation to realistic quark
masses should be quite smooth. It is therefore feasible to include the light
quarks with current lattice technology. Once these systematic effects have
been removed, simulations should be accurate to a few percent, where
nonperturbative contributions to the couplings may become important.

Simulations using the techniques presented in this paper should, in the near
future, produce accurate spectra, decay rates, wavefunctions,
and other matrix elements for all
of the lowest-lying mesons in the $\psi$ and $\Upsilon$ families. These
techniques should also be useful in studies of $D$ and $B$, as NRQCD
propagators are more efficiently generated than relativistic propagators,
and much less afflicted by noise than static propagators \cite{lepmac}.
With these methods we are entering a new era of
high-precision lattice simulations of quantum chromodynamics.

\section*{Acknowledgements}
We are grateful to C.~Davies, I.~Drummond, R.~Horgan, P.~Mackenzie, and
B.~Thacker for discussions, comments, and suggestions. This work was supported
in part by a grant from the National Science Foundation.
K.~H. was supported in part by the Department of Energy and by Ohio State
University.
U.~M. was supported by a scholarship from the Academy of Finland, and would
like
to thank Cornell University for hospitality during the completion of this work.

\vskip 1in

\noindent \appendix{{\Large {\bf Appendix}}}

\label{app}

\vskip 0.3cm

This appendix discusses the construction in NRQCD of lattice operators
which reproduce the corresponding continuum operators to any desired
order in $a$, at tree level.  We first describe a natural technique
which works for all bilinears in fermion fields, but turns out to
be inconvenient in actual simulations.  Then we show how to circumvent
this problem by starting with a convenient lattice operator
and computing corrections to it.  Finally, we discuss in some detail
the corrections to the gluon action in the nonabelian case.

Consider a general local, gauge covariant, continuum operator, bilinear in the
fermion fields and with canonical dimension $d = n + 3$,
 \beq
 O(x) = \psid (x) K_n (\Ev,\Bv,\Dv,D_t) \psi (x) \, .
 \eeq
Since the field strength components are proportional to commutators of
covariant derivatives, such an operator is a linear combination of the tensors
 \beq
 \label{Dtensor}
 O_{\mu_1, \ldots ,\mu_n} (x) \equiv \psid (x) D_{\mu_1}
         \ldots D_{\mu_n} \psi (x) \, .
 \eeq
It is straightforward to construct lattice expressions for these tensors,
accurate
to a specific order in $a$, provided one defines the link variables correctly
in
terms of the continuum gauge potentials, as in \eq{link}. The relationship
between
the continuum covariant derivatives and the corresponding finite differences on
the lattice is then given by \eq{exactdef}, and one can rewrite the tensors in
\eq{Dtensor} using, for example, the symmetric expression
 \beq
 \label{centder}
 D_\mu = \frac{1}{2 a} \left[ \ln (1 + a \Delta^{(+)}_\mu) -
                         \ln (1 - a \Delta^{(-)}_\mu) \right] \, .
 \eeq
In practice of course one always works with the first few terms in the power
series
expansion of \eq{centder}. One can then use the fact that $\Delta^{(+)}_\mu =
\Delta^{(-)}_\mu + \O(a)$ to construct an expression for $D_\mu$ that is
maximally
local on the lattice, as for example \eq{imprD}.
To see what kind of lattice expressions are generated through this procedure,
consider for example
 \beq
 O_{[\mu \nu]} (x) \equiv - \, i g \, \psid (x) F_{\mu \nu} (x) \psi(x) \, .
 \eeq
Using \eq{imprD}, this becomes, up to corrections ${\cal O}(a^4)$,
 \beqa
 O_{[\mu \nu]} (x) & = & \psid (x) \left[ \left( \, \Delta^{(\pm)}_\mu
             \Delta^{(\pm)}_\nu \right. \right. - \frac{a^2}{6}
              \Delta^{(\pm)}_\mu \Delta^{(+)}_\nu \Delta^{(\pm)}_\nu
               \Delta^{(-)}_\nu \\
   & - & \left. \left. \frac{a^2}{6} \Delta^{(+)}_\mu \Delta^{(\pm)}_\mu
         \Delta^{(-)}_\mu \Delta^{(\pm)}_\nu \right)
          - \left( \mu \Leftrightarrow \nu \right) \right] \psi (x) \nonumber
\, .
 \eeqa
In terms of link variables the result is simpler and more transparent, and it
is
shown in Fig. 2. There one can see the main characteristic common
to all lattice operators constructed this way: the quark field and its
hermitian
conjugate are evaluated at different points on the lattice.
In actual simulations, this turns out to be a serious disadvantage:
it is much more economical to define both fermions at the same site.

Fortunately, it is as easy to go from a lattice operator written
in terms of link variables to the corresponding continuum expression.
We can therefore start with any operator having the appropriate
continuum limit, such as our cloverleaf $F_{\mu\nu}^{(c)} (x)$,
and then correct it to the desired order in $a$.  Consider
for example the first term in the cloverleaf definition of the field strength
tensor, sandwiched between spinors:
 \beq
 \label{firstquad}
 \psid (x) U_{x,\mu} U_{x + a \hat{\mu}, \nu} U^{\dagger}_{x + a \hat{\nu},
\mu}
     U^{\dagger}_{x,\nu} \psi (x) \, .
 \eeq
It is easy to find an expression written in terms of lattice derivatives that
will reproduce this term. One starts by replacing each matrix $U_\mu$ with a
forward derivative $\Delta^{(+)}_\mu$, and each matrix $U^\dagger_\mu$ with the
corresponding backward derivative. This gives
 \beq
 \psid (x) \Delta^{(+)}_\mu \Delta^{(+)}_\nu \Delta^{(-)}_\mu
              \Delta^{(-)}_\nu \psi (x) \, ,
 \eeq
which contains \eq{firstquad}, plus correction terms with at most three
$U$'s, that are easily calculated. The procedure is then iterated until all
terms with a lower number of $U$'s are canceled. Once the desired lattice
operator
is written in terms of lattice derivatives, the corresponding continuum
operator
can be calculated using, once again, \eq{exactdef}, to the desired order
in $a$.

This procedure identifies the terms which need to be subtracted from the
original definition.  Once known in the continuum, it is easy
to guess a lattice expression that will cancel them, yielding the
corrected lattice operator.  The result for the cloverleaf field strength
is, as expected, given by \eq{corrclov} and \eq{newcorrclov}.

The correction to the gluon action in the nonabelian theory is somewhat more
complicated to check directly.
We start by noting that,
to all orders in $a$, the link variable can be written as
 \beq
 U_{x,\mu} = \exp \left[ \sum_{n = 1}^\infty \left( a^n u_{n,\mu}^b T_b \right)
              \right] \, .
 \eeq
This is a consequence of the Baker-Campbell-Hausdorff theorem,
that allows to rewrite the path-ordered
exponential as exponential of a series of commutators. Our definition of the
link
variable, with the gauge potentials evaluated at mid-link, implies for example
 \beqa
 \label{linkterms}
 u_{1,\mu}^b & = & - i g A^b_\mu \nl
 u_{2,\mu}^b & = & u_{4,\mu}^b \; = \; 0 \\
 u_{3,\mu}^b = - i \frac{g}{24} \partial_\mu^2 A_\mu^b \! \! & - & \! \!
               i \frac{g^2}{12} f_{b c d} A_\mu^c \partial_\mu A_\mu^d \, ,
 \nonumber
 \eeqa
where $f_{b c d}$ are the structure constants of $SU(3)$.
By the same token, we can write the plaquette matrix as
 \beq
 \label{plaqgen}
 U_{P(x,\mu \nu)} = \exp \left[ \sum_{n = 1}^\infty \left( a^n C_{n,\mu \nu}^b
T_b
                      \right) \right] \, ,
 \eeq
where, with the fields at the center, one easily finds that
 \beq
 \label{c1c2}
 C_{1,\mu \nu}^b = 0 \; \; \; C_{2,\mu \nu}^b = - i g F_{\mu \nu}^b \, .
 \eeq
The series expansion in powers of $a$ of the trace of \eq{plaqgen} can be
considerably simplified by using the trace identities
 \beqa
 {\rm Tr}(T_a) = 0 \; & \; & \; {\rm Tr}(T_a T_b) = \frac{1}{2} \delta_{a b}
\nl
 {\rm Tr}(T_a T_b T_c) \! & = & \! \frac{1}{4} (d_{a b c} + i f_{a b c}) \, ,
 \eeqa
where $d_{a b c}$ is the completely symmetric invariant tensor of $SU(3)$.
This yields
 \beqa
 \label{trplaqC}
 {\rm Tr} \left( U_{P(x,\mu \nu)} - 1 \right)
      & = & \frac{a^4}{4} \Cv_{2,\mu \nu} \cdot
      \Cv_{2,\mu \nu} + \frac{a^5}{2} \Cv_{2,\mu \nu} \cdot \Cv_{3,\mu \nu} \nl
    & + & \frac{a^6}{2} \Cv_{2,\mu \nu} \cdot \Cv_{4,\mu \nu} +
          \frac{a^6}{4} \Cv_{3,\mu \nu} \cdot \Cv_{3,\mu \nu} \\
    & + & \frac{a^6}{24} C_{2,\mu \nu}^a C_{2,\mu \nu}^b C_{2,\mu \nu}^c
            d_{a b c} + \O(a^7) \, , \nonumber
 \eeqa
so that the only coefficients we need to calculate are $\Cv_{3,\mu \nu}$ and
$\Cv_{4,\mu \nu}$.
The calculation is rather lengthy, but
can be performed using a symbolic manipulation program, such as
``Mathematica'' \cite{wolfram}. The result is
 \beqa
 \label{c3c4}
 C_{3,\mu \nu}^a T_a & = & - \frac{g^2}{2} \left[ A_\mu + A_\nu, F_{\mu \nu}
        \right]  \nl
 C_{4,\mu \nu}^a T_a & = & - \frac{i g }{24} (D_\mu^2 + D_\nu^2) F_{\mu \nu}
\nl
       & + & \frac{g^2}{8} \left[ (\partial_\mu + \partial_\nu) (A_\mu +
A_\nu),
                F_{\mu \nu} \right] \\
       & + & \frac{i g^3}{8} \left[ A_\mu + A_\nu, \left[ A_\mu + A_\nu,
                F_{\mu \nu} \right] \right] \, . \nonumber
 \eeqa
Substituting \eq{c1c2} and \eq{c3c4} in \eq{trplaqC} one can verify, as a
nontrivial check, that all gauge non-invariant terms disappear, as do the
coefficients of odd powers of $a$. The final result is, as expected
 \beq
 {\rm Tr} \left[ R \left( U_{P(x,\mu \nu)} \right) \right] =
 - \frac{g^2 a^4}{2} {\rm Tr}(F_{\mu \nu}^2)
 - \frac{g^2 a^6}{24} {\rm Tr} \left[ F_{\mu \nu} (D_\mu^2 + D_\nu^2)
          F_{\mu \nu} \right] \, ,
 \eeq
the nonabelian generalization of \eq{reoneplaq}.

\clearpage


\begin{table}[h]

\begin{center}

 \begin{tabular}{c|c|l}
 Operator& Estimate      & Description \\ \hline
 \rule{0cm}{.5cm}
 $\psi$  &  $(Mv)^{3/2}$ & quark (annihilation) field \\
 $\chi$  &  $(Mv)^{3/2}$ & antiquark (creation) field \\
 $D_t$   &  $M v^2$      & gauge covariant time derivative \\
 $\Dv$   &  $M v$        & gauge covariant spatial derivative \\
 $g\phi$ &  $M v^2$      & scalar potential (Coulomb gauge) \\
 $g\Av$  &  $M v^3$      & vector potential (Coulomb gauge) \\
 $g\Ev$  &  $M^2v^3$     & chromoelectric field \\
 $g\Bv$  &  $M^2v^4$     & chromomagnetic field \\
 \end{tabular}

 \caption{The component fields and operators for the NRQCD action for heavy
quarks. The estimated magnitude of each in a quarkonium state
and in Coulomb gauge is
given in terms of the quark mass $M$ and typical velocity $v$.}

\label{fields}

\end{center}

\end{table}

\begin{table}[h]

\label{mftest}

\begin{center}

 \begin{tabular}{rc|c|c}
 & $Ma$ & P.Th. & M.F.Th. \\ \cline{2-4}\cline{2-4} \rule{0cm}{.5cm}
$a E_0/g^2$: & $\infty$ & 0.17 & 0.08 \\
          & 5 & 0.21 & 0.14 \\
          & 2.5 & 0.25 & 0.18 \\
          & 1.5 & 0.30 & 0.23 \\ \cline{2-4}
$(Z_M-1)/g^2$: & $\infty$ & 0.07 & 0.08 \\
          & 5 & 0.04 & 0.03 \\
          & 2.5 & 0.07 & 0.03 \\
          & 1.5 & 0.06 & 0.0 \\ \cline{2-4}
$(Z_\psi-1)/g^2$: & $\infty$ & 0.04 & 0 \\
          & 5 & 0.07 & 0.05 \\
          & 2.5 & 0.10 & 0.10 \\
          & 1.5 & 0.15 & 0.17 \\
 \end{tabular}

 \caption{The coefficient of $g^2$ in expansions of the NRQCD
renormalization parameters. Results coming from exact perturbation
theory (P.Th.) are compared with estimates based on mean-field theory
(M.F.Th.) for a variety of masses~$M$. The parameters are appropriate to the
evolution equation used by Davies and Thacker.
For $Z_M$, $n=1$ for $Ma=\infty$ and $5$;
$n=2$ for $Ma=2.5$ and $1.5$.  $E_0$ and $Z_\psi$ are
independent of $n$.  An infrared divergence,
which is shared with the continuum theory, has been
omitted from $Z_\psi$.}

 \end{center}

 \end{table}

\end{document}